\documentclass[aps,prb,showpacs,psfig]{revtex4}
\usepackage{amsfonts,psfig} 

\begin{document}

\title{Parametric resonance of a two-dimensional electron gas under 
bichromatic irradiation}
\author{C.\ Joas}
\affiliation {Fachbereich Physik, Freie Universit\"at Berlin, 
Arnimallee 
14, D-14195 Berlin, Germany}
\author {M.E.\ Raikh}
\affiliation{Department of Physics, University of Utah, Salt Lake City, 
UT 84112}
\author{F.\ von Oppen}
\affiliation{Fachbereich Physik, Freie Universit\"at Berlin, 
Arnimallee 
14, D-14195 Berlin,
Germany\footnote{Permanent address}\\
Dept.\ of Condensed Matter Physics, 
Weizmann Institute of Science, Rehovot 76100, Israel}
 
\date{\today}

\begin{abstract}
  In an ultrahigh mobility 2D electron gas, even a weak
  nonparabolicity of the electron dispersion, by violating Kohn's
  theorem, can have a drastic effect on {\it dc} magnetotransport
  under {\it ac} drive. In this paper, we study theoretically the
  manifestation of this effect in the {\it dc} response to the
  combined action of {\em two} driving {\it ac}-fields (bichromatic
  irradiation).  Compared to the case of monochromatic irradiation,
  which is currently intensively studied both experimentally and
  theoretically, the presence of a second microwave source provides
  additional insight into the properties of an {\it ac}-driven 2D
  electron gas in weak magnetic field.  In particular, we find that
  nonparabolicity, being the simplest cause for a violation of Kohn's
  theorem, gives rise to new qualitative effects specific to
  bichromatic irradiation.  Namely, when the frequencies $\omega_1$
  and $\omega_2$ are well away from the cyclotron frequency,
  $\omega_c$, our simple classical considerations demonstrate that the
  system becomes unstable with respect to fluctuations with frequency
  $\frac{1}{2}\left(\omega_1+\omega_2\right)$.  The most favorable
  condition for this {\it parametric} instability is
  $\frac{1}{2}\left(\omega_1+\omega_2\right)\simeq\omega_c$.  The
  saturation level of this instability is also determined by the
  nonparabolicity.  We also demonstrate that, as an additional effect
  of nonparabolicity, this parametric instability can manifest itself
  in the {\it dc} properties of the system. This happens when
  $\omega_1$, $\omega_2$ and $\omega_c$ are related as $3:1:2$,
  respectively.  Even for weak detuning between $\omega_1$ and
  $\omega_2$, the effect of the bichromatic irradiation on the {\it
    dc} response in the presence of nonparabolicity can differ
  dramatically from the monochromatic case.  In particular, we
  demonstrate that, beyond a critical intensity of the two fields, the
  equations of motion acquire {\em multistable} solutions. As a
  result, the diagonal {\it dc}-conductivity can assume {\em several}
  stable negative values at the {\em same} magnetic field.

\end{abstract}
\pacs{05.60.-k,73.43.Cd,73.50.Pz}

\maketitle

\section{Introduction}

The cyclotron resonance in a 2D electron gas was first studied almost
30 years ago.\cite{kotthaus75,abstreiter76} These studies revealed an
oscillatory magnetoabsorption of microwave radiation with its
principal peak at $\omega = \omega_c$, where $\omega_c$ and $\omega$
are the cyclotron and microwave frequencies, respectively, and several
subharmonics at $\omega = n\omega_c$ due to the disorder-induced
violation of Kohn's theorem.

Recently, interest in the properties of microwave-driven electrons in
a magnetic field has been revived,\cite{zudov01} especially after
experiments\cite{mani02,zudov03} carried out on samples with extremely
high mobilities, indicated that near the cyclotron resonance and its
harmonics, irradiation results in drastic changes of the diagonal $dc$
resistivity.  In contrast, the Hall resistivity remains practically
unchanged by illumination, and retains its classical value. The
experimental observations\cite{mani02,zudov03} were confirmed in a
number of subsequent studies.
\cite{yang03,mani03,dorozhkin03,willett03,studenikin03,zudov04,mani04,
  mani'04,studenikin04} This unusual behavior of the weak-field
magnetoconductivity is currently accounted for by an instability
resulting from a sign reversal of the diagonal photoconductivity under
irradiation.\cite{andreev03,bergeret03,shi03} The developed
instability results in dynamical symmetry breaking,\cite{andreev03}
i.e., in an inhomogeneous state of the system characterized by domains
of current flowing in opposite directions.  Ongoing theoretical
studies
\cite{durst03,lei03,ryzhii03,ryzhii03',dmitriev03,vavilov04,dmitriev03'}
concentrate on the microscopic description of the sign reversal of
photoconductivity. Closely related physics was already discussed
theoretically quite long ago.\cite{ryzhii70,ryzhii86}

Obviously, a complete understanding of the fascinating properties of
ultraclean 2D electron systems under irradiation requires additional
experimental studies. At the same time, the number of feasible
measurements that were not carried out so far is limited.  A promising
avenue seems to be to study the response to {\it bichromatic}
irradiation. Motivated by this, in the present paper we calculate this
response within the simple model\cite{koulakov03} of a clean {\em
  classical}\cite{Kovalev04} 2D gas, in which Kohn's theorem is
violated due to nonparabolicity
\begin{equation}
\varepsilon(p)= \frac{p^2}{2m}\left[1 -
\frac{p^2}{2mE_0}\right],
\label{disp}
\end{equation}
of the electron dispersion.  Here $m$ is the effective mass, and $E_0$
is an energy of the order of the bandgap.

Denote by ${\cal E}_1$ and ${\cal E}_2$ the amplitudes of two linearly
polarized {\it ac} fields with frequencies $\omega_1$ and $\omega_2$,
respectively. In the presence of a {\it dc} field $E_{dc}$, the
equation of motion for the electron momentum ${\cal P} = p_x + ip_y$
takes the form
\begin{equation}
\label{motion}
\frac{d {\cal P}}{d t} + \frac {{\cal P}}{\tau} - i \omega_c
{\cal P} +
\frac{i \omega_c}{m E_0} {\cal P} |{\cal P}|^2 = e
E_{dc}e^{i\theta}+
\frac{{e{\cal E}_1}}{2} \Bigl( e^{i\omega_1 t}+ e^{-i\omega_1
t}\Bigr)
+ \frac{{e{\cal E}_2}}{2} \Bigl( e^{i\omega_2 t}+ e^{-i\omega_2
t}\Bigr),
\end{equation}
where $\omega_c$ is the cyclotron frequency, $\tau$ is the relaxation
time, and $\theta$ is the orientation of the weak {\it dc} field with
respect to the fields ${\cal E}_1$, ${\cal E}_2$, which we assume to
be parallel to each other.  For a monochromatic {\em ac} drive, ${\cal
  E}_2 =0$, it was demonstrated \cite{koulakov03} that within a
certain interval of magnetic fields near the cyclotron resonance, Eq.\ 
(\ref{motion}) yields negative diagonal conductivity, $\sigma_d <0$,
without significant change of the Hall conductivity. This sign
reversal occurs when the mobility is high, $\omega_c\tau \gg 1$, and
${\cal E}_1$ is sufficiently strong. In the vicinity of the cyclotron
resonance, $\sigma_d$ turns negative even when the irradiation-induced
change of the electron mass is relatively weak. Thus, the simple model
Eq.\ (\ref{motion}) exhibits negative photoconductivity without
invoking Landau quantization. It also predicts bistable hysteretic
behavior of $\sigma_d$ as a function of the detuning from the
cyclotron resonance for large enough $\omega_c\tau$.

\begin{figure}[t]
\begin{center}
$\begin{array}{ccccc} 
\psfig{file=fig1a.eps,width=1.4in}&\quad &
\psfig{file=fig1b.eps,width=1.4in}&\quad &
\psfig{file=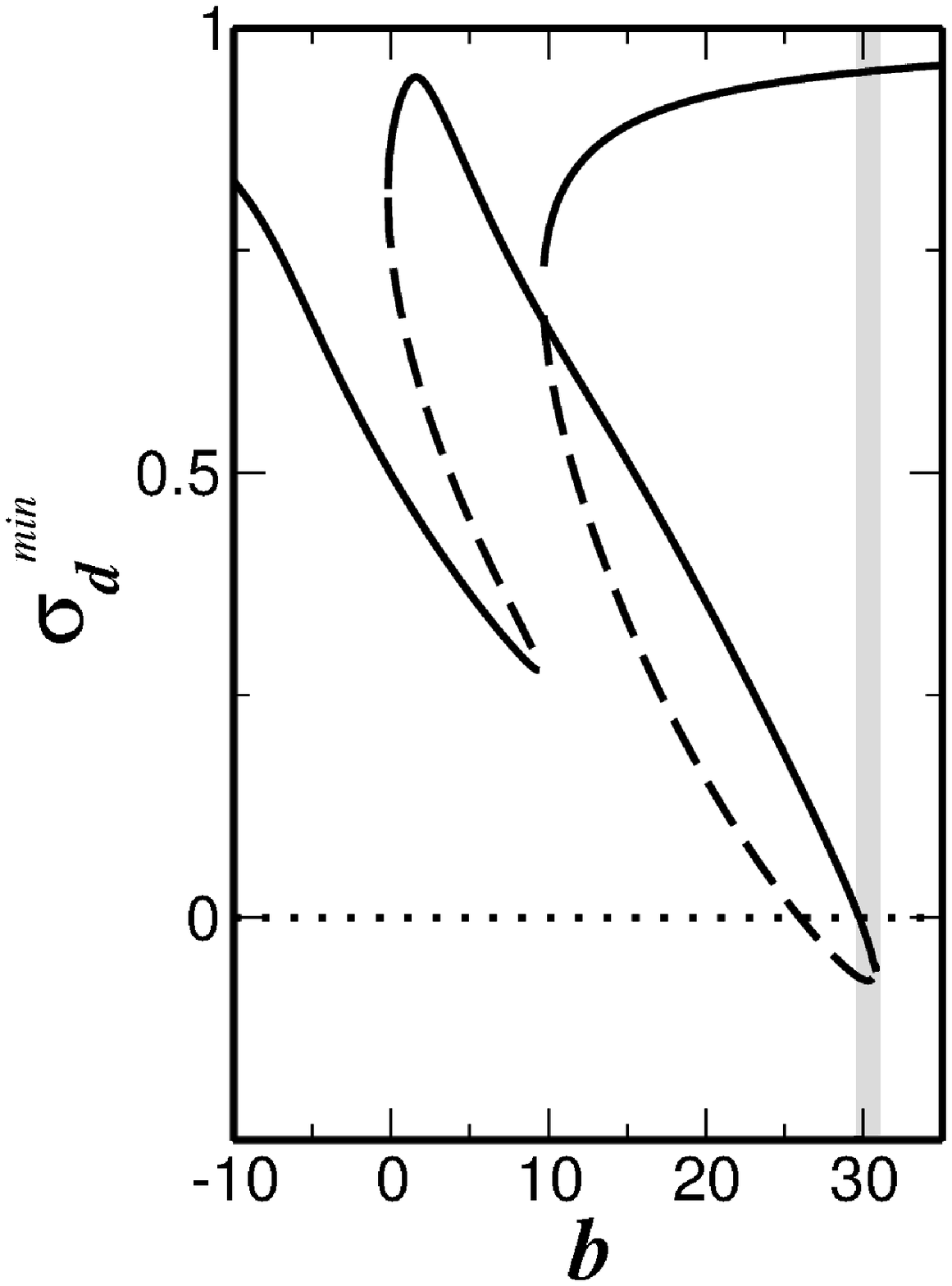,width=1.4in}\\
\mbox{(a)}&& \mbox{(b)} && \mbox{(c)}
\end{array}$
\end{center}
\caption{Dimensionless diagonal conductivity (in units of the Drude value) plotted from Eq.\ (\ref{14}) versus the dimensionless magnetic field, defined by Eq.\ (\ref{xdef}), for three cases: (a) {\em monochromatic} irradiation with frequency $\omega_1$ and dimensionless intensity $A=14.4$; (b)   {\em monochromatic} irradiation with the same intensity as in (a) and frequency $\omega_2=5\omega_1/3$; (c) {\em bichromatic} case: response to simultaneous irradiation with two microwave sources having the same intensities and frequencies as in (a) and (b). The emerging region of negative diagonal conductivity is shaded. 
All three plots (a)-(c) are calculated for $\frac{1}{2}\left(\omega_1+\omega_2\right)\tau=20$. Full and dashed lines correspond to stable and unstable branches, respectively.}
\label{Fig0}
\end{figure}

In the present paper, we extend the consideration of Ref.\ 
\onlinecite{koulakov03} to the bichromatic case.  The most convincing
illustration that the response to irradiation with two {\it ac}-fields
cannot simply be reduced to the superposition of the responses to each
individual field, is presented in Fig.\ \ref{Fig0}.  It is seen in
Figs.\ \ref{Fig0}(a,b) that the individual fields of equal intensity
and frequency ratio $5:3$ are unable to reverse the sign of the
diagonal conductivity at any magnetic field.  At the same time, upon
{\em simultaneous} irradiation by the both fields, a domain of
magnetic fields emerges, within which the diagonal conductivity is
negative (see Fig.\ \ref{Fig0}(c)).

In addition, our study reveals the following new features that are
specific to the bichromatic case:

\noindent ({\em i}) The presence of the second {\em ac} field 
on the r.h.s.\ of Eq.\ (\ref{motion}) gives rise to a second domain of
magnetic field, within which $\sigma_d$ is negative.  Upon increasing
the intensities of the two {\em ac} fields, the two domains of
negative photoconductivity merge into a single domain which broadens
much faster with the {\em ac} intensity than in the monochromatic
case.

\noindent ({\em ii}) For monochromatic irradiation,
$\sigma_d$ could assume either one or two stable values. By contrast,
under bichromatic irradiation, we find a {\em multistable} regime
within certain domains of magnetic field.

\noindent({\em iii}) In the vicinity of the conditions 
$(\omega_1+\omega_2)=2\omega_c$ and $\vert \omega_1-\omega_2\vert =
2\omega_c$, a nonparabolicity-induced {\it parametric} instability
develops in the system.  As a result of this instability, the
components $(\omega_1+\omega_2)/2$ and $\vert
\omega_1-\omega_2\vert/2$ emerge in addition to the conventional
frequencies $\omega_1$ and $\omega_2$ of the momentum oscillations.
These components, in turn, upon mixing with the components $\omega_1$,
$\omega_2$, give rise to components of ${\cal P}$ oscillating with
frequencies $3\omega_1-\omega_2$. Thus, for {\em bichromatic}
irradiation, two high-frequency {\em ac} driving fields can create a
low frequency current circulating in the system. In particular, for
$\omega_2= 3\omega_1 \approx 3\omega_c/2$ the system exhibits a {\em
  dc} response to the {\em ac} drive.

The paper is organized as follows. The case of weak detuning from the
cyclotron frequency is treated analytically in Sec.\ \ref{weak}. In
the same section, we present numerical results in this limit, which
exhibit nontrivial multistable behavior.  In Sec.\ \ref{strong}, we
consider the case of strong detuning, where we find a
nonparabolicity-induced parametric resonance.  Concluding remarks are
presented in Sec. \ref{conclusions}.

\section{Weak detuning}
\label{weak}

For monochromatic irradiation, the cyclotron resonance develops when
the microwave frequency is close to the cyclotron frequency
$\omega_c$. In this section, we consider bichromatic irradiation when
both frequencies $\omega_1$ and $\omega_2$ are close to $\omega_c$,
$\vert\omega_1-\omega_c\vert \ll \omega_c$ and
$\vert\omega_2-\omega_c\vert \ll \omega_c$, so that the cyclotron
resonances due to $\omega_1$ and $\omega_2$ can interfere with one
another.

\subsection{Calculation of diagonal conductivity}

In analogy to Ref.\ \onlinecite{koulakov03}, we search for solutions
of Eq.\ (\ref{motion}) in the form
\begin{equation}
\label{complex}
{\cal P}(t)={\cal P}_0+ {\cal P}_{1}^{+}\exp(i\omega_1t) +
{\cal P}_{1}^{-}\exp(-i\omega_1 t)+{\cal P}_{2}^{+}\exp(i\omega_2t)+
{\cal P}_{2}^{-}\exp(-i\omega_2 t),
\end{equation}
where ${\cal P}_0$ is a small {\em dc} component proportional to
$E_{dc}$. The components ${\cal P}_{1}^{-}$ and ${\cal P}_{2}^{-}$ are
nonresonant and can be found from the simplified equations
\begin{equation}
\label{simp1}
-i(\omega_1+\omega_c){\cal P}_{1}^{-} = \frac{e{\cal E}_1}{2},
\end{equation}
\begin{equation}
\label{simp2}
-i(\omega_2+\omega_c){\cal P}_{2}^{-} = \frac{e{\cal E}_2}{2},
\end{equation}
where we neglect both relaxation and nonlinearity. However, relaxation
and nonlinearity must be taken into account when calculating the
resonant components ${\cal P}_{1}^{+}$ and ${\cal P}_{2}^{+}$.
Substituting Eq.\ (\ref{complex}) into Eq.\ (\ref{motion}), and taking
into account that $|{\cal P}_{1}^{-}|,|{\cal P}_{2}^{-}|\ll |{\cal
  P}_{1}^{+}|, |{\cal P}_{2}^{+}|$, we arrive at a system of coupled
equations for the resonant momentum components,
\begin{equation}
\left[i\left(\omega_1-\omega_c\right)+\frac{1}{\tau}+\frac{i\omega_c}{m
E_0}\left(\vert{\cal P}_1^{+} \vert^2+ 2 \vert{\cal
P}_2^{+}\vert^2\right)\right]{\cal P}_1^{+}=\frac{e{\cal E}_1}{2} ,
\label{P1eq}
\end{equation}
\begin{equation}
\left[i\left(\omega_2-\omega_c\right)+\frac{1}{\tau}+
\frac{i\omega_c}{m E_0}\left(2 \vert{\cal P}_1^{+}\vert^2+  
\vert{\cal P}_2^{+}\vert^2\right)\right]{\cal P}_2^{+}=
\frac{e{\cal E}_2}{2}.
\label{P2eq}
\end{equation}
Despite the inequalities $\vert{\cal P}_1^{+}\vert\ll\vert{\cal
  P}_1^{-}\vert$ and $\vert{\cal P}_2^{+}\vert\ll\vert{\cal
  P}_2^{-}\vert$, it is crucial to keep the nonresonant components
${\cal P}_1^{-}$ and ${\cal P}_2^{-}$ when considering the {\em dc}
component ${\cal P}_0$. This yields
\begin{equation}
\left[-i\omega_c +\frac{1}{\tau}+\frac{2i\omega_c}{m
E_0}\left(\vert{\cal P}_1^{+} \vert^2+\vert{\cal
P}_2^{+}\vert^2\right)\right]{\cal P}_0+\frac{2i\omega_c}{mE_0}
\left[{\cal P}_1^{+}{\cal P}_1^{-}+ {\cal P}_2^{+}{\cal P}_2^{-}\right]{\cal P}_0 ^*
= eE_{dc}e^{i\theta}.
\label{P0eq}
\end{equation}
Due to the nonlinearity, the microwave intensities induce an effective
shift in the resonance frequency $\omega_c$. Thus, it is convenient to
introduce effective detunings $\Omega_1$ and $\Omega_2$ by
\begin{eqnarray}
\label{effective1}
\Omega_1&=&\omega_1-\omega_c +
\frac{\omega_c}{mE_0}\left(\vert{\cal P}_1^{+} \vert^2+ 2 \vert{\cal
P}_2^{+}\vert^2\right),\\
\label{effective2}
\Omega_2&=&\omega_2-\omega_c +
\frac{\omega_c}{mE_0}\left(2\vert{\cal P}_1^{+} \vert^2+  \vert{\cal
P}_2^{+}\vert^2\right),
\end{eqnarray}
and to present formal solutions of Eqs.\ (\ref{P1eq}) and (\ref{P2eq})
in the form
\begin{equation}
\label{formal}
{\cal P}_1^{+}= \frac{e{\cal E}_1\tau}{2\left(1+i\Omega_1\tau\right)} \quad,\quad  \quad \quad{\cal P}_2^{+}= \frac{e{\cal E}_2\tau}{2\left(1+i\Omega_2\tau\right)}.
\end{equation}
Note that the detunings $\Omega_1$ and $\Omega_2$ themselves depend on
${\cal P}_1^{+}$ and ${\cal P}_2^{+}$, so that Eqs.\ 
(\ref{effective1}), (\ref{effective2}) and (\ref{formal}) should be
considered as a system of nonlinear equations for the resonant
momentum components ${\cal P}_1^{+}$ and ${\cal P}_2^{+}$.  Assuming
that the detunings $\Omega_1$ and $\Omega_2$ are known, the solution
of Eq.\ (\ref{P0eq}) yields for the $dc$ component
\begin{equation}
\label{solution}
{\cal P}_0=\frac{eE_{dc}}{\omega_c^2\tau}
\Biggl[\left(1+i\omega_c\tau\right)e^{i\theta}+
\frac{1}{4mE_0}\left\{\frac{\left(e{\cal E}_1\tau\right)^2}
{1+i\Omega_1\tau} +\frac{\left(e{\cal E}_2\tau\right)^2}
{1+i\Omega_2\tau}\right\}e^{-i\theta}\Biggr].
\end{equation} 
The diagonal conductivity $\sigma_d$ is proportional to ${\rm
  Re}\left({\cal P}_0e^{-i\theta}\right)$.  Thus, the second term in
Eq.\ (\ref{solution}) gives rise to a $\theta$-dependence of the
nonparabolicity-induced contribution to the diagonal conductivity
which is given by $\sin\left(2\theta - \phi\right)$. Here, $\phi$
satisfies the equation
\begin{equation}
\label{phi}
\tan\phi = \frac{\Omega_1\tau\left(1+\Omega_2^2\tau^2\right){\cal E}_1^2
+\Omega_2\tau\left(1+\Omega_1^2\tau^2\right){\cal E}_2^2}
{\left(1+\Omega_2^2\tau^2\right){\cal E}_1^2 +
\left(1+\Omega_1^2\tau^2\right){\cal E}_2^2}.
\end{equation}
We deduce that the minimal value of $\sigma_d$ is given by
\begin{equation}
\sigma_d^{min}=\frac{ne^2}{m\omega_c^2\tau}\Biggl\{1-
\frac{e^2\tau^2}{4mE_0}
\left[\frac{\left({\cal E}_1^2\Omega_2\tau+
{\cal E}_2^2\Omega_1\tau\right)^2+
\left({\cal E}_1^2+{\cal E}_2^2 \right)^2}
{\left(1+\Omega_1^2\tau^2\right)\left(1+\Omega_2^2\tau^2\right)}\right]
^{1/2}
\Biggr\}.
\label{14}
\end{equation}
In the following, we will be particularly interested in
$\sigma_d^{min}$, since the condition $\sigma_d^{min}<0$ is sufficient
for the formation of the zero-resistance state.

\subsection{Numerical results: Multistability}

As demonstrated in Ref. \onlinecite{koulakov03}, the diagonal
conductivity in the monochromatic case shows a region of bistability.
In the bichromatic case under study, there can even be multistable
behavior as will now be shown.  We measure the frequency difference of
the {\it ac}-fields by $\Delta=\left(\omega_1-\omega_2\right)\tau$ and
the magnetic field by
\begin{equation}
b=\left(\omega_c-\frac{\omega_1+\omega_2}{2}\right)\tau
\label{xdef},
\end{equation}
which depends linearly on the magnetic field $B$.  Upon substituting
the formal solutions ${\cal P}_1 ^{+}$ and ${\cal P}_2 ^{+}$ of Eq.\ 
(\ref{formal}) into Eqs.\ (\ref{effective1}-\ref{effective2}), these
can be written as a pair of coupled equations for the effective
detunings $\Omega_1\tau$ and $\Omega_2\tau$
\begin{equation}
\Omega_1\tau=\frac{\Delta}{2}-b+A\left[\frac{1}{1+\left(\Omega_1\tau\right)^2}+\frac{2\eta^2}{1+\left(\Omega_2\tau\right)^2}\right]
,\quad\quad
\Omega_2\tau=-\frac{\Delta}{2}-b+A\left[\frac{2}{1+\left(\Omega_1\tau\right)^2}+\frac{\eta^2}{1+\left(\Omega_2\tau\right)^2}\right],
\label{system}
\end{equation}
where $\eta$ and $A$, given by
\begin{equation}
\eta={\cal E}_2/{\cal E}_1,\quad \quad \quad A=\omega_c\tau\frac{\left(e{\cal E}_1\tau\right)^2}{4mE_0},
\label{etaandA}
\end{equation}
measure the ratio of the field amplitudes and the ratio of the
absolute field intensities to the nonparabolicity of the electron
spectrum, respectively. As in the monochromatic case, the strength of
the first order correction to the Drude conductivity is proportional
to the microwave intensity and thus $A$. At a fixed magnetic field $b$
and at fixed {\it ac} frequencies and amplitudes, this coupled system
of two third-order equations can yield up to nine simultaneous
solutions $(\Omega_1\tau,\Omega_2\tau)$ for the effective detunings.
Since $\sigma_d ^{min}$ is directly related to these effective
detunings via
\begin{equation}
\sigma_d ^{min} =\sigma_{D}\left\{ 1-\frac{A}{\omega_c\tau}\left[\frac{1}{1+\left(\Omega_1\tau\right)^2}+\frac{\eta^4}{1+\left(\Omega_2\tau\right)^2}+\frac{2\eta^2\left(1+\Omega_1\tau\Omega_2\tau\right)}{\left(1+\left(\Omega_1\tau\right)^2\right)\left(1+\left(\Omega_2\tau\right)^2\right)}\right]^{1/2}\right\},
\label{sigmadminimal}
\end{equation}
where $\sigma_{D}=ne^2/(\omega_c^2\tau)$ is the Drude conductivity,
there are thus up to nine individual branches of $\sigma_d ^{min}$ at
a given $b$. This multistable behavior occurs in the vicinity of the
resonance and will be studied below for the specific case of $\eta=1$.

\begin{figure}[t]
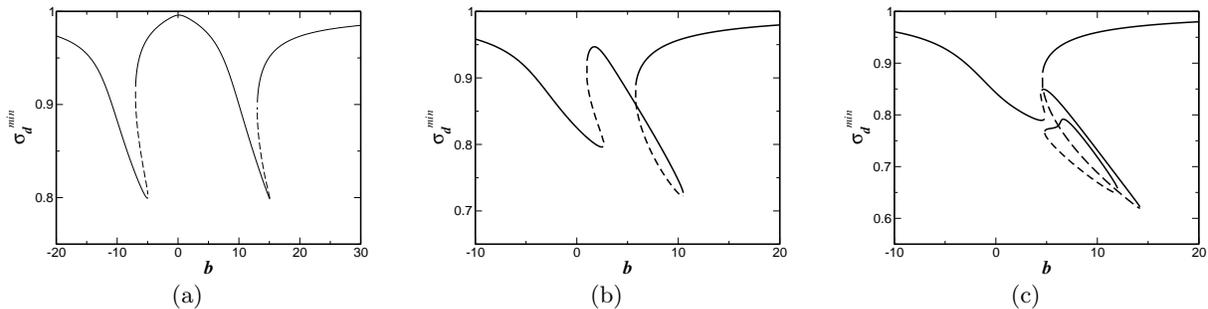

\begin{center}
$\begin{array}{ccccc}
\psfig{file=fig2a.eps,width=1.9in}&\quad &
\psfig{file=fig2b.eps,width=1.9in} & \quad &
\psfig{file=fig2c.eps,width=1.9in}\\
\mbox{(a)}&& \mbox{(b)} && \mbox{(c)}
\end{array}$
\end{center}
\caption{Evolution of the dimensionless  (in units of the Drude conductivity $\sigma_D$) minimal conductivity $\sigma_d^{min}$ as function of magnetic field $b$, defined in Eq.\ (\ref{xdef}), for three different values of $\Delta$; (a) $\Delta=20$, (b) $\Delta=5$, (c) $\Delta=1$. The curves are calculated for the values of parameters $\eta=1$,  $A=5$ (defined by Eq.\ (\ref{etaandA})), and $\omega_c\tau=25$. The distance of the two dips that can be clearly discerned in (a) is roughly $\Delta$. When lowering $\Delta$, the dips move closer together (b) and finally merge (c). In addition, multistable regions emerge. As in Fig.\ \ref{Fig0}, unstable branches are plotted as dashed lines.}
\label{Fig1}
\end{figure}

We first focus on the dependence of $\sigma_d ^{min}(b)$ on $\Delta$.
For large $\Delta$, i.e. markedly different {\it ac}-frequencies, we
expect two separate regions in $b$ where $\sigma_d ^{min}$ deviates
significantly from the Drude result. These are the regions where the
cyclotron frequency is in resonance with one of the two {\it ac}
frequencies, i.e. either $\omega_1\simeq\omega_c$ or
$\omega_2\simeq\omega_c$. Inside these regions, the behavior with
respect to $b$ is very similar to the monochromatic case, except that
the irradiation-induced effective shift of $\omega_c$ now depends on
both external frequencies.  In particular, the emergence of bistable
regions inside these two separate intervals as in the monochromatic
case is to be expected. This can be seen in Fig.\ \ref{Fig1}(a), where
$\sigma_d ^{min}$ is shown as a function of magnetic field $b$ for
rather large $\Delta$. Two dips in $\sigma_d ^{min}$ can be clearly
discerned, the inner branches of which are unstable.  Upon reducing
$\Delta$, the two dips move closer together up to a point where the
frequencies $\omega_1$ and $\omega_2$ are so close that the analogy to
the monochromatic case breaks down and the two dips start to interact
to finally from a single multistable dip in the limit
$\Delta\rightarrow 0$. This behavior is exemplified in Figs.\ 
\ref{Fig1}(b,c). It can be seen that multiple solutions of $\sigma_d
^{min}(b)$ develop upon reducing $\Delta$.

\begin{figure}[b]
\begin{center}
$\begin{array}{ccccc}
\psfig{file=fig3a.eps,width=1.9in}&\quad &
\psfig{file=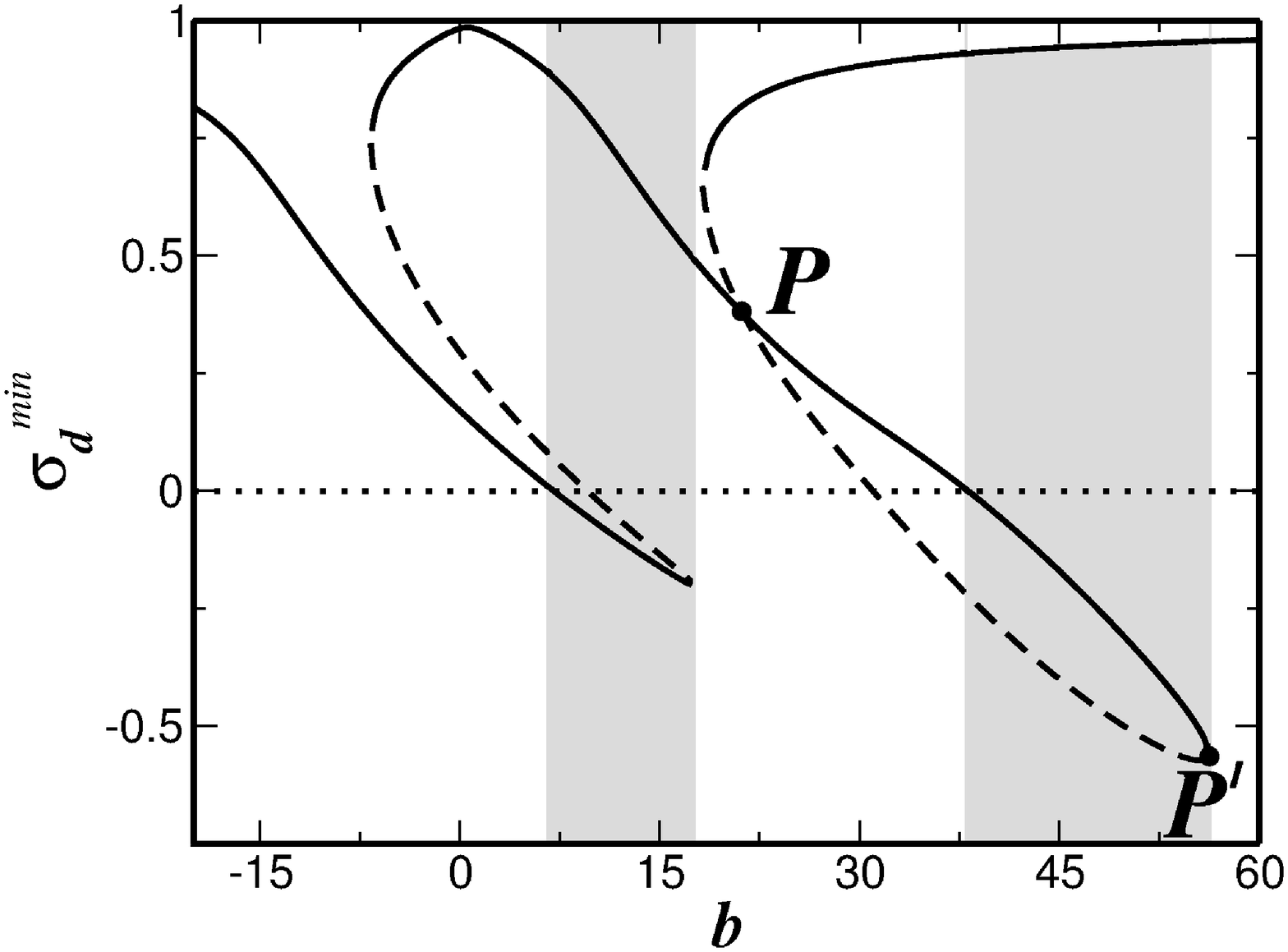,width=1.96in}&\quad&
\psfig{file=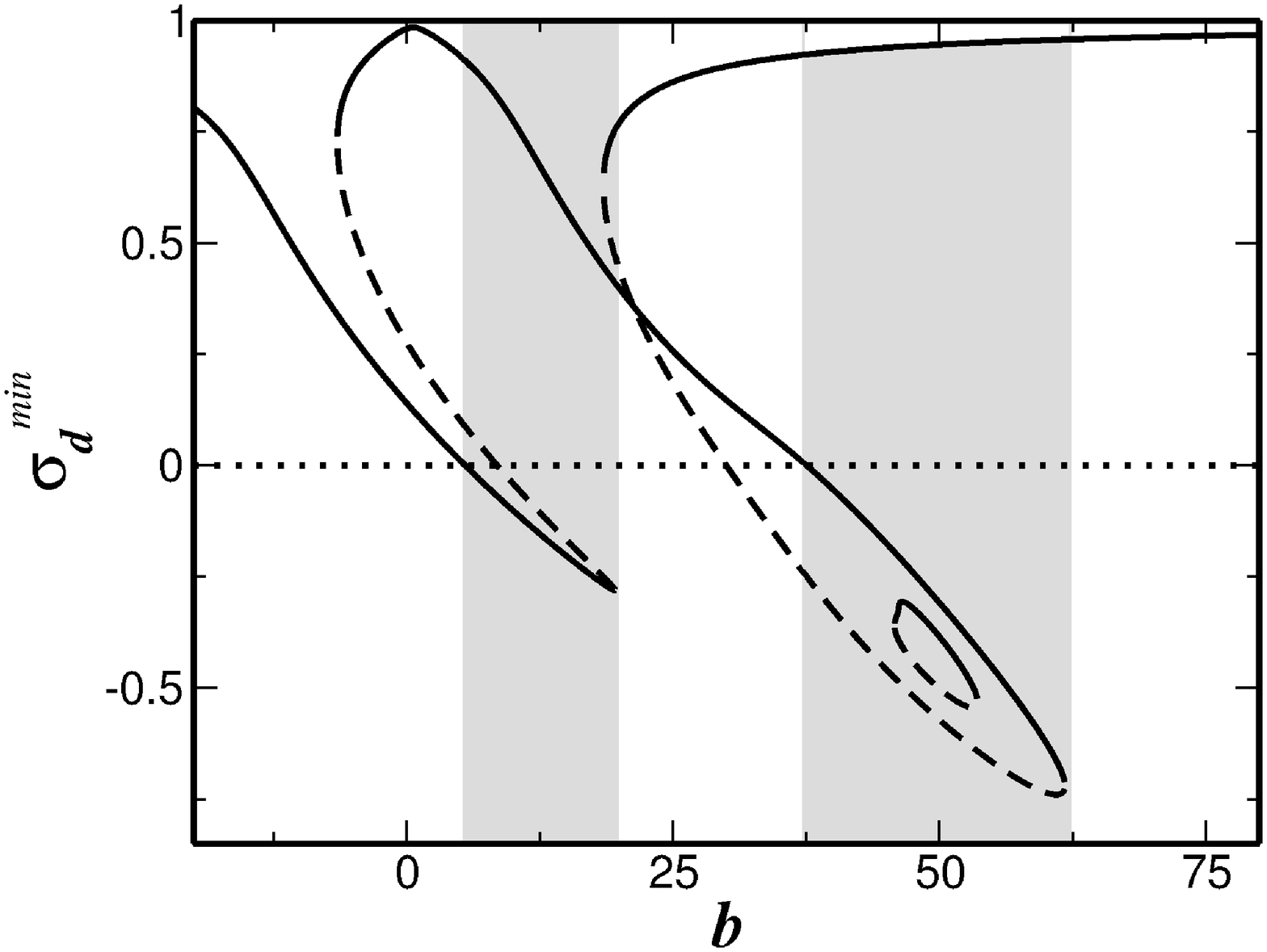,width=1.9in}\\
\mbox{(a)}&& \mbox{(b)} && \mbox{(c)}
\end{array}$
\end{center}
\caption{The dimensionless minimal diagonal conductivity, $\sigma_d^{min}(b)$, defined by Eq.\ (\ref{sigmadminimal}), is plotted for three different values of the dimensionless intensity $A$ of the two {\it ac}-fields; (a) $A=20$, (b) $A=30$, (c) $A=32$. The domains of magnetic field $b$ with negative $\sigma_d^{min}$ are shaded. The dotted line indicates the boundary between positive and negative $\sigma_d^{min}$. Unstable branches are dashed as in Fig.\ \ref{Fig0}. The curves are calculated for the values of the parameters $\eta=1$, $\Delta=(\omega_1-\omega_2)\tau=25$ and $\omega_c\tau=25$. In (b), the point ${\bf P}$ is a point where a continuous stable and a continuous unstable branch intersect without ``noticing'' each other, while the point ${\bf P^\prime}$ is a cusp which separates a stable from an unstable branch. }
\label{Fig2}
\end{figure}

Next, we consider the case of negative diagonal conductivity and study
the evolution of $\sigma_d ^{min}$ with magnetic field.  As expected,
there is a threshold intensity below which no negative diagonal
conductivity is observed. Upon increasing the field amplitudes and
thus $A$, the negative first order correction to the Drude
conductivity grows linearly with $A/(\omega_c\tau)$ as can be seen
from Eq.\ (\ref{sigmadminimal}). When this correction exceeds one,
negative $\sigma_d ^{min}$ is to be expected in some regions of
magnetic field $b$. Fig.\ \ref{Fig2} shows the $b$-dependence of
$\sigma_d ^{min}$ for three specific values of $A$. It can be seen
that at low $A$, no regions in $b$ with negative diagonal conductivity
can be observed. At higher $A$, two regions in $b$ show negative
$\sigma_d ^{min}$-branches as is also indicated by the shaded regions
in Fig.\ \ref{Fig2}(b). For even higher $A$, a single large region in
$b$ shows negative diagonal conductivity.

\begin{figure}[tttt]
\begin{center}
\centerline{\psfig{file=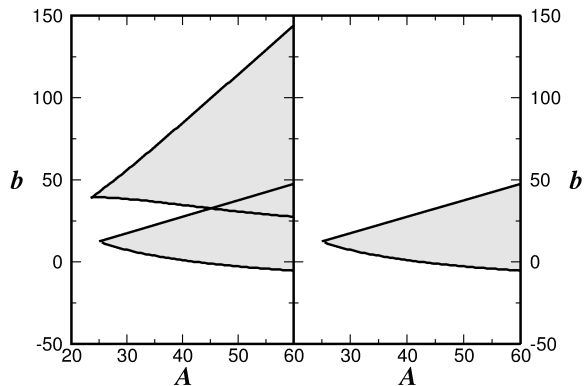,width=3in}}
\end{center}
\caption{Evolution of the regions of negative $\sigma_d^{min}$ with irradiation intensity, $A$. Shown are the bichromatic case (left panel) and the monochromatic case (right panel). In both cases, only the dominant (stable) branches are shown to avoid confusion. The parameters are the same as in Fig.\ \ref{Fig2}.}
\label{Fig3}
\end{figure}

To clarify the evolution of these regions with increasing field
amplitudes, we plotted the extension of the regions in $b$ as a
function of $A$. The result is shown in Fig.\ \ref{Fig3}. It is
remarkable that above the threshold value of $A$, first a single
region appears that shows negative $\sigma_d ^{min}$. Then, in the
immediate vicinity of the threshold a second, well separated region
develops. Upon further increasing $A$, the width of these regions
grows and, eventually, the two regions merge to form a single broad
region of negative diagonal conductivity. For comparison, we also show
the monochromatic case in the right hand panel of Fig.\ \ref{Fig3}.

\subsection{Stability of different branches}

The stability of the various branches of $\sigma_d^{min}(b)$ as shown
by solid and dashed lines in Figs.\ \ref{Fig0},\ref{Fig1}, and
\ref{Fig2}, can be obtained from a standard stability analysis.  As
usual, we find that stable and unstable branches ``meet" at cusps, as
clearly seen at the minima of $\sigma_d$ in Figs.\ \ref{Fig0} and
\ref{Fig1}. The transitions from unstable to stable branches at larger
$b$ in these figures are also accompanied by cusps, although this can
not necessarily be discerned within the resolution of the figures. The
number of branches increases with the irradiation intensity, cf.\ 
Figs.\ \ref{Fig1} and \ref{Fig2}.  The rule that stable and unstable
branches meet in cusps remains valid, although this statement becomes
less trivial. For example, in Fig.\ \ref{Fig2}(b) stable and unstable
branches intersect at point ${\bf P}$ without ``noticing each other".
Accordingly, there is no cusp at this point. At the same time, there
is a cusp at ${\bf P^\prime}$ in Fig.\ \ref{Fig2}(b) where the same
branches switch between stable and unstable. Figs.\ \ref{Fig1}(c) and
\ref{Fig2}(c) illustrate how new branches and multistability emerge
with increasing irradiation intensity. The emergence of new stable and
unstable branches occurs in pairs which meet at additional cusps.  In
both figures \ref{Fig1}(c) and \ref{Fig2}(c), there are regions in
magnetic field with three coexisting stable solutions (tristability).
Further increase of $A$ would lead to up to eight cusps in Fig.\
\ref{Fig2}(c), each of which is a meeting point of stable and unstable
branches.  Thus, the tristability situation illustrated in Fig.\ 
\ref{Fig2}(c) will evolve into a magnetic field domain with
``four-stability''.

\section{Strong detuning}
\label{strong}

In this section, we consider the case when both frequencies $\omega_1$
and $\omega_2$ are tuned away from $\omega_c$.  This implies that the
system (\ref{P1eq}-\ref{P2eq}) decouples and acquires the obvious
solutions
\begin{equation}
{\cal P}_1^{+}=\frac{e{\cal E}_1\tau}
{2\left[1+i\left(\omega_1-\omega_c\right)\tau\right]}, ~~~~~~~~~~~\quad 
{\cal P}_2^{+}=\frac{e{\cal E}_2\tau}{2\left[1+i\left(\omega_2-\omega_c\right)
\tau\right]}.
\label{P1P2}
\end{equation}
The condition $e{\cal E}_1$, $e{\cal E}_2 \ll \omega_c
\left(mE_0\right)^{1/2}$ for decoupling follows from Eqs.\ 
(\ref{P1eq}-\ref{P2eq}), assuming that $\vert\omega_1-\omega_c\vert$,
$\vert\omega_2-\omega_c\vert$ $\sim \omega_c\gg 1/\tau$.
Interestingly, even under this condition, the solutions are unstable
for certain relations between the frequencies $\omega_1$, $\omega_2$.
The mechanism for this instability relies on mixing of the two
external drive frequencies by the nonparabolicity which results in a
modulation of the effective cyclotron frequency. This modulation, in
turn, can lead to {\it parametric} resonance.

To perform the stability analysis of the solutions Eq.\ (\ref{P1P2}),
we introduce a small deviation ${\cal P} \rightarrow {\cal P} + \delta
{\cal P}$ and linearize Eq.\ (\ref{motion}) with respect to $\delta
{\cal P}$. The linearized equation (\ref{motion}) has the form
\begin{equation}
\label{linearized}
\frac{d}{dt}\left(\delta {\cal P}\right)+ \Biggl(\frac{1}{\tau}
-i\omega_c+\frac{2i\omega_c}{mE_0}\vert{\cal P}\vert^2\Biggr)
\delta {\cal P}
+\frac{i\omega_c}{mE_0}{\cal P}^2\left(\delta {\cal P}\right)^{\ast}=0.
\end{equation}
This equation couples $\delta {\cal P}$ to $\left(\delta {\cal
    P}\right)^{\ast}$ via the nonparabolicity of the electron
spectrum.  The corresponding equation for $\delta {\cal P}^{\ast}$
reads
\begin{equation}
\label{conjugate}
\frac{d}{dt}\left(\delta {\cal P}^{\ast}\right)+ \Biggl(\frac{1}{\tau}
+i\omega_c - \frac{2i\omega_c}{mE_0}\vert{\cal P}\vert^2\Biggr)
\delta {\cal P}^{\ast} - \frac{i\omega_c}{mE_0}
\left({\cal P}^{\ast}\right)^2\delta {\cal P}=0.
\end{equation}
The coupling coefficient, ${\cal P}^2$, as seen from Eq.\ 
(\ref{complex}), contains the harmonics $\pm 2\omega_1$, $\pm
2\omega_2$, $\pm (\omega_1+\omega_2)$, and $\pm (\omega_1-\omega_2)$.
This suggests that $\delta{\cal P}(t)$ also contains a number of
harmonics, namely, $\pm \omega_1$, $\pm \omega_2$, $\pm
(\omega_1+\omega_2)/2$, and $\pm (\omega_1-\omega_2)/2$. An
instability might develop when one of these frequencies is close to
$\omega_c$. Thus, in the monochromatic case, the instability develops
only in the vicinity of the cyclotron resonance $\omega_1 \approx
\omega_c$. The branches, shown with dashed lines, in Figs.\ 
\ref{Fig0}(a,b), are unstable due to this instability. By contrast,
the bichromatic case offers two additional options for an instability
to develop, even if the frequencies $\omega_1$, $\omega_2$ are
nonresonant, namely $\omega_c \approx
\left(\omega_1+\omega_2\right)/2$ and $\omega_c \approx
\vert\left(\omega_1-\omega_2\right)\vert/2$.  The considerations of
both cases are analogous to each other.  Therefore, we focus on the
first case below.

\subsection{Parametric instability at 
  $\left(\omega_1+\omega_2\right)\approx 2\omega_c$}

Upon substituting the ansatz
\begin{equation}
\label{form}
\delta{\cal P}(t)=C\exp\left\{\left[\Gamma + 
\frac{i\left(\omega_1+\omega_2\right)}{2}\right]t\right\},~~~~~~
\delta{\cal P}(t)^{\ast}=C^{\ast}\exp\left\{\left[\Gamma - 
\frac{i\left(\omega_1+\omega_2\right)}{2}\right]t\right\}
\end{equation}
into Eqs.\ (\ref{linearized}-\ref{conjugate}) and keeping only
resonant terms, we obtain the following system of algebraic equations
for $C$ and $C^{\ast}$
\begin{eqnarray}
\label{following1}
\Biggl[\Gamma + \frac{1}{\tau}+
\frac{i\left(\omega_1+\omega_2-2\omega_c\right)}{2}+
\frac{2i\omega_c}{mE_0}\left(\vert{\cal P}_1^{+}\vert^2 +
\vert{\cal P}_2^{+}\vert^2\right)\Biggr]C &=& -\frac{2i\omega_c}{mE_0}
{\cal P}_1^{+}{\cal P}_2^{+}C^{\ast} \\
\label{following2}
\Biggl[\Gamma + \frac{1}{\tau}-
\frac{i\left(\omega_1+\omega_2-2\omega_c\right)}{2}-
\frac{2i\omega_c}{mE_0}\left(\vert{\cal P}_1^{+}\vert^2 +
\vert{\cal P}_2^{+}\vert^2\right)\Biggr]C^{\ast} &=& 
\frac{2i\omega_c}{mE_0}
\left({\cal P}_1^{+}{\cal P}_2^{+}\right)^{\ast}C.
\end{eqnarray}  
Thus, the most favorable condition for instability is determined by
the following relation between $\omega_1$ and $\omega_2$
\begin{eqnarray}
\label{favorable}
\omega_1+\omega_2 &=& 2\omega_c\left[1-\frac{2}{mE_0}
\left(\vert{\cal P}_1^{+}\vert^2 +
\vert{\cal P}_2^{+}\vert^2\right) \right]\nonumber \\
&\approx&  
2\omega_c\left\{1-\frac{e^2}{2mE_0}\left[\frac{{\cal E}_1^2}
{\left(\omega_1-\omega_c\right)^2} +\frac{{\cal E}_2^2}
{\left(\omega_2-\omega_c\right)^2}\right]\right\}
\approx
2\omega_c\left[1 -\frac{2e^2\left({\cal E}_1^2 + {\cal E}_2^2\right)}
{mE_0\left(\omega_1-\omega_2\right)^2}\right],
\end{eqnarray}
where Eq.\ (\ref{P1P2}) has been used.  In this case, the increment
$\Gamma$ is maximal and given by
\begin{equation}
\label{increment}
\Gamma_{max}= -\frac{1}{\tau}+\left(\frac{2\omega_c}{mE_0}\right)
\vert{\cal P}_1^{+}{\cal P}_2^{+}\vert 
\approx -\frac{1}{\tau}+ 
\left\vert\frac{e^2\omega_c{\cal E}_1{\cal E}_2}{mE_0
\left(\omega_1-\omega_c\right)\left(\omega_2-\omega_c\right)}\right\vert
\approx  -\frac{1}{\tau}+ 
\frac{e^2\left(\omega_1+\omega_2\right)\vert{\cal E}_1{\cal E}_2\vert}
{mE_0\left(\omega_1-\omega_2\right)^2}.
\end{equation}
The parametric instability develops if $\Gamma_{max}$ is positive.  It
is important to note that the condition $\Gamma_{max}>0$ is consistent
with the condition of strong detuning when the simplified expressions
Eq.\ (\ref{P1P2}) are valid. Indeed, assuming
$\vert\omega_1-\omega_c\vert \sim \vert\omega_2-\omega_c\vert \sim
\omega_c$, the two conditions can be presented as
$\omega_c\left(mE_0\right)^{1/2} \gg e{\cal E}_1, e{\cal E}_2 \gg
\tau^{-1}\left(mE_0\right)^{1/2}$. Therefore, for $\omega_c\tau \gg
1$, there exists an interval of the amplitudes of the {\em ac} fields
within which both conditions are met.  Note also, that parametric
resonance does not develop exactly at $\omega_c =
(\omega_1+\omega_2)/2$, i.e. at $b=0$ (in dimensionless units, see
Eq.\ (\ref{xdef})). In fact, from Eqs.\ (\ref{favorable}) and
(\ref{increment}) it can be concluded that $\Gamma_{max}>0$
corresponds to $b \gtrsim 1$. In experimental situations, when
$\omega_1$ and $\omega_2$ are fixed, Eq.\ (\ref{favorable}) can also
be viewed as an expression for the magnetic field
$\omega_c=\omega_c^{opt}$, at which the parametric instability is most
pronounced.  The interval of $\omega_c$ around $\omega_c^{opt}$,
within which the increment is positive can be found from the
dependence $\Gamma\left(\omega_c\right)$
\begin{equation}
\label{interval1}
\Gamma(\omega_c)= -\frac{1}{\tau}+
\sqrt{\left(\Gamma_{max}+\frac{1}{\tau}\right)^2-
\Bigl(\omega_c-\omega_c^{opt}\Bigr)^2}\approx -\frac{1}{\tau}+
\sqrt{\left[\frac{e^2\left(\omega_1+\omega_2\right)
\vert{\cal E}_1{\cal E}_2\vert}
{mE_0\left(\omega_1-\omega_2\right)^2}\right]^2-
\Bigl(\omega_c-\omega_c^{opt}\Bigr)^2}.
\end{equation}
Upon setting $\Gamma(\omega_c)=0$ in the l.h.s.  of Eq.\ 
(\ref{interval1}), we find the width of the interval to be
\begin{equation}
\label{interval2}
\vert\omega_c-\omega_c^{opt}\vert \leq 
\left\{\left[\frac{e^2\omega_c\vert{\cal E}_1{\cal E}_2\vert}
{2mE_0\left(\omega_1-\omega_c\right)^2}\right]^2 -~
\frac{1}{\tau^2}\right\}^{1/2}\approx~ 
\left\{\left[\frac{e^2\left(\omega_1+\omega_2\right)
\vert{\cal E}_1{\cal E}_2\vert}
{mE_0\left(\omega_1-\omega_2\right)^2}\right]^2 -~
\frac{1}{\tau^2}\right\}^{1/2}.
\end{equation}

It is instructive to reformulate the condition for the parametric
instability in a different way. Assume for simplicity that ${\cal
  E}_1={\cal E}_2$. Then the combination $e^2\vert{\cal E}_1{\cal
  E}_2\vert/2mE_0(\omega_1-\omega_c)^2$ is equal to $\delta m/m$,
where $\delta m/m$ is the relative correction to the electron
effective mass due to irradiation.\cite{koulakov03} From Eq.\ 
(\ref{increment}) it follows that the condition $\Gamma_{max}> 0$ can
be presented as $\left(\omega_c\tau\right)\left(\delta m/m\right)> 1$.
With $\omega_c\tau \gg 1$ this condition can be satisfied even for
$\delta m \ll m$. Note, that in the case of weak detuning, the same
condition is required for the {\em dc} conductivity to assume negative
values near the cyclotron resonance.

Summarizing, we arrive at the following scenario. In the case of
strong detuning, there is no mutual influence of the responses to the
{\em ac} fields ${\cal E}_1$ and ${\cal E}_2$ as long as they are
weak. However, as the product $\vert{\cal E}_1{\cal E}_2\vert$
increases and reaches a critical value $\vert{\cal E}_1{\cal
  E}_2\vert_c$, the threshold, $\Gamma_{max}=0$, where $\Gamma_{max}$
is given by Eq.\ (\ref{increment}), is exceeded at the magnetic field
$\omega_c = \omega_c^{opt}$ determined by Eq.\ (\ref{favorable}).
Above the threshold, fluctuations with frequencies close to
$\left(\omega_1+\omega_2\right)/2$ are amplified. This effect of
parametric instability is solely due to nonparabolicity.  Then the
natural question arises: At what level does the component of momentum
with frequency $\left(\omega_1+\omega_2\right)/2$ saturate above
threshold? This question is addressed in the next subsection.

\subsection{Parametric instability at  
  $\vert\omega_1-\omega_2\vert\approx 2\omega_c$}

We now briefly discuss parametric instability at weak magnetic field,
$\omega_c \approx \vert\left(\omega_1-\omega_2\right)/2\vert$.  Assume
for concreteness, that $\omega_1 > \omega_2$. In this case, the
optimal magnetic field, $\tilde\omega_c^{opt}$, is lower and reads
\begin{equation}
\label{low}
\omega_1-\omega_2 \approx  
2\tilde\omega_c^{opt}\left\{1-\frac{e^2}{2mE_0}\left[\frac{{\cal E}_1^2}
{\left(\omega_1-\tilde\omega_c^{opt}\right)^2} +\frac{{\cal E}_2^2}
{\left(\omega_2-\tilde\omega_c^{opt}\right)^2}\right]\right\}
\approx 2\tilde\omega_c^{opt}
\left\{1-\frac{2e^2}{mE_0}\left[\frac{{\cal E}_1^2}
{\left(\omega_1+\omega_2\right)^2} +\frac{{\cal E}_2^2}
{\left(3\omega_2-\omega_1\right)^2}\right]\right\}.
\end{equation}
At $\omega_c = \tilde\omega_c^{opt}$, the threshold condition for
parametric instability, analogous to Eq.\ (\ref{increment}), has the
form
\begin{equation}
\label{tilde}
\tilde\Gamma_{max} \approx  -\frac{1}{\tau}+ 
\frac{e^2\left(\omega_1-\omega_2\right)\vert{\cal E}_1{\cal E}_2\vert}
{mE_0\left(\omega_1+\omega_2\right)\vert 3\omega_2-\omega_1\vert} > 0.
\end{equation}
There is no real divergence in Eqs.\ (\ref{low}), (\ref{tilde}) in the
limit $\omega_1 \rightarrow 3\omega_2$, since these are derived under
the assumption that the difference $\vert 3\omega_2-\omega_1\vert$ is
$\gtrsim 1/\tau$.

\subsection{Saturation of parametric resonance}

As the threshold for parametric resonance is exceeded, the harmonics
with frequency $\left(\omega_1+\omega_2\right)/2$ can no longer be
considered as a perturbation, but rather have to be included into the
equation of motion. In other words, we must search for a solution of
Eq.\ (\ref{motion}) in the form
\begin{equation}
{\cal P}=  {\cal P}_1^{+}\exp\left(i\omega_1 t \right)+ 
{\cal P}_2^{+} \exp\left(i\omega_2 t \right) + 
{\cal P}_3(t) \exp\left[i\left(\frac{\omega_1+\omega_2}{2}\right)t 
\right],
\end{equation}
where ${\cal P}_3(t)$ is a slowly varying function of time.
Upon substituting this form into Eq.\ (\ref{motion}), 
we obtain the following coupled equations for ${\cal P}_3(t)$
and ${\cal P}^{\ast}_3(t)$
\begin{equation}
\label{coupled1}
\frac{d{\cal P}_3}{dt}+ \Biggl[\frac{1}{\tau}+
\frac{i\left(\omega_1+\omega_2-2\omega_c\right)}{2}+
\frac{i\omega_c}{mE_0}\left(2\vert{\cal P}_1^{+}\vert^2 +
2\vert{\cal P}_2^{+}\vert^2 +\vert{\cal P}_3\vert^2\right)\Biggr]
{\cal P}_3 = -\frac{2i\omega_c}{mE_0}
{\cal P}_1^{+}{\cal P}_2^{+}{\cal P}_3^{\ast},
\end{equation}      
\begin{equation}
\label{coupled2}
\frac{d{\cal P}^{\ast}_3}{dt}+\Biggl[\frac{1}{\tau}-
\frac{i\left(\omega_1+\omega_2-2\omega_c\right)}{2}-
\frac{i\omega_c}{mE_0}\left(2\vert{\cal P}_1^{+}\vert^2 +
2\vert{\cal P}_2^{+}\vert^2+\vert{\cal P}_3\vert^2\right)\Biggr]
{\cal P}^{\ast}_3 = 
\frac{2i\omega_c}{mE_0}
\left({\cal P}_1^{+}{\cal P}_2^{+}\right)^{\ast}{\cal P}_3.
\end{equation}
Saturated parametric instability is described by setting $d{\cal
  P}_3/dt =0$ and $d{\cal P}^{\ast}_3/dt =0$ in Eqs.\ (\ref{coupled1})
and (\ref{coupled2}), respectively.  The result for ${\cal P}_3$ has
the simplest form for the optimal magnetic field $\omega_c =
\omega_c^{opt}$
\begin{equation}
\label{satur}
\vert{\cal P}_3\vert(\omega_c^{opt}) = 
\Biggl[\vert{\cal P}_1^{+}{\cal P}_2^{+}\vert^2
-\frac{4m^2E_0^2}{\left(\omega_1+\omega_2\right)^2\tau^2}
\Biggr]^{1/4}.
\end{equation}
From Eq.\ (\ref{satur}) we conclude that, in the vicinity of the
threshold, $\vert{\cal P}_3\vert$ increases as $\Bigl(\vert{\cal
  E}_1{\cal E}_2\vert - \vert{\cal E}_1{\cal
  E}_2\vert_c\Bigr)^{1/4}\propto\Gamma_{max}^{1/4}$.  Well above the
threshold it approaches the value $\sqrt{\vert{\cal P}_1^{+}{\cal
    P}_2^{+}\vert}$. From here we conclude that, even upon saturation,
the magnitude of the nonparabolicity-induced harmonics with frequency
$\left(\omega_1+\omega_2\right)/2$ does not have a ``back'' effect on
the magnitudes Eq.\ (\ref{P1P2}) of the responses to the {\em ac}
fields.

For magnetic fields in the vicinity of $\omega_c^{opt}$ the saturation
value, $\vert{\cal P}_3\vert (\omega_c)$, is given by
\begin{equation}
\label{satur1}
\vert{\cal P}_3\vert (\omega_c)= 
\Biggl\{\Biggl[\vert{\cal P}_1^{+}{\cal P}_2^{+}\vert^2
-\frac{4m^2E_0^2}{\left(\omega_1+\omega_2\right)^2\tau^2}
\Biggr]^{1/2}-\frac{2mE_0\vert\omega_c-\omega_c^{opt}\vert}
{\left(\omega_1+\omega_2\right)}\Biggr\}^{1/2}\simeq
\left[\vert{\cal P}_3\vert^2(\omega_c^{opt})-
\frac{2mE_0\vert\omega_c-\omega_c^{opt}\vert}
{\left(\omega_1+\omega_2\right)}\right]^{1/2}.
\end{equation}
In contrast to $\vert{\cal P}_3\vert(\omega_c^{opt})$, the threshold
behavior of $\vert{\cal P}_3\vert (\omega_c)$ is slower, namely
$\vert{\cal P}_3\vert (\omega_c)\propto \Bigl(\vert{\cal E}_1{\cal
  E}_2\vert - \vert{\cal E}_1{\cal E}_2\vert_c\Bigr)^{1/2}$.  In
principle, one has to verify that the solutions Eqs.\ (\ref{satur}),
(\ref{satur1}), that describe the saturated parametric resonance, are
stable. This can be done with the use of the system Eqs.\ 
(\ref{coupled1}), (\ref{coupled2}), by perturbing it around the
saturated solution. The outcome of this consideration is that the
corresponding perturbations do indeed decay.

\subsection{Implications for {\em dc} transport}

As we demonstrated in the previous subsection, the parametric
instability that develops in the case of two {\em ac} fields above a
certain threshold, results in the component of the momentum, ${\cal
  P}_3\exp\left[i\left(\omega_1+\omega_2\right)t/2\right]$, which well
above the threshold has the magnitude $\vert{\cal P}_3\vert\approx
\sqrt{\vert{\cal P}_1\vert\vert{\cal P}_2\vert}$.  The important
consequence of the developed parametric resonance is, that the
component ${\cal P}_3$ gives rise to new harmonics in the term
$\propto \vert{\cal P}\vert^2 {\cal P}$ in the equation of motion Eq.\ 
(\ref{motion}). Of particular interest are the ${\cal P}_3$-induced
terms
\begin{equation}
\left[{\cal P}_1^2{\cal P}_3^{\ast}+{\cal P}_1{\cal P}_2^{\ast}
{\cal P}_3\right]
\exp\left[i\frac{3\omega_1-\omega_2}{2}t\right].
\end{equation}
It is easy to see that, under the condition $\omega_2\approx
3\omega_1$, these terms act as an effective {\em dc} field, and thus
generate low-frequency {\em circular} current even without {\em dc}
drive. If the relation between the frequencies is precisely $1:3$,
then the magnetic field, at which the developed parametric instability
would give rise to a quasistationary circular current distribution,
can be determined from Eq.\ (\ref{favorable})
\begin{equation}
\omega_c\simeq2\omega_1\left[1 + \frac{2e^2\left({\cal E}_1^2 + {\cal E}_2^2\right)}
{mE_0\left(\omega_1-\omega_2\right)^2}\right].
\end{equation}
If the ratio $\omega_2/\omega_1$ is close, but not exactly $1:3$,
there is still a certain allowance, determined by Eq.\ 
(\ref{interval2}) for the formation of the quasistationary current.
The above effect of spontaneous formation of {\em dc}-like currents
under irradiation is distinctively different from the formation of
current domains when $\sigma_d$ turns negative under irradiation.
Firstly, the effect is specific to bichromatic irradiation. Secondly,
it requires rather strict commensurability between the two
frequencies, and finally, it develops within a very narrow interval
around a certain magnetic field.

\section{Conclusions}
\label{conclusions}

In the present paper we have considered the problem of single electron
motion in a magnetic field under irradiation by two monochromatic
fields. When the frequencies $\omega_1$ and $\omega_2$ differ only
slightly, $\vert\omega_1-\omega_2\vert \sim 1/\tau$, the effect of a
weak nonparabolicity of the electron spectrum on the diagonal
conductivity is qualitatively the same for
monochromatic\cite{koulakov03} and bichromatic irradiation.  The prime
qualitative effect which distinguishes the bichromatic case is the
emergence of a parametric resonance at magnetic fields
$\omega_c=(\omega_1+\omega_2)/2$ and $\omega_c=\vert
\omega_1-\omega_2\vert/2$ when the detuning is strong (of the order of
the cyclotron frequency).  It is instructive to compare this effect
with the parametric resonance of electrons in a magnetic field due to
a weak time modulation of the field amplitude.\cite{Aronov1,
  Aronov2,Aronov3} The latter effect, considered more than 20 years
ago, has a transparent explanation.  The modulation of the magnitude
of a $dc$ field with frequency $2\omega_c$ translates into a
corresponding modulation of the cyclotron frequency, so that the
equation of motion of the electron reduces to that for a harmonic
oscillator with a weakly time-modulated eigenfrequency. The solution
of this equation is unstable, if the modulation frequency is close to
$2\omega_c$.  As a natural stabilizing mechanism of the parametric
resonance, the authors of Ref.\ \onlinecite{Aronov1} considered the
nonparabolicity Eq.\ (\ref{disp}) of the electron dispersion.  For a
characteristic magnetic field of $B=0.1$ T, the cyclotron frequency is
$\omega_c=3.6\cdot10^{11}$ Hz, i.e.\ in the microwave range so that
conventional modulation of $B$ with a frequency $2 \omega_c$ is
technically impossible. To bypass this obstacle, it was proposed in
Ref.\ \onlinecite{Aronov3} to use microwave illumination with
frequency $2 \omega_c$ to create a parametric resonance. The idea was
that the {\it magnetic field} component of the pumping electromagnetic
wave would provide the necessary oscillatory correction to the
external {\it dc} magnetic field.  In the present paper, we have
demonstrated that two nonresonant {\it ac} sources can {\it enforce} a
parametric resonance of the type considered in Refs.\ 
\onlinecite{Aronov1,Aronov2,Aronov3} {\it without} any time modulation
of the {\it dc} magnetic field. Remarkably, this
bichromatic-radiation-induced cyclotron resonance emerges due to the
same nonparabolicity Eq.\ (\ref{disp}) that played the role of a
stabilizing factor in Refs.\ \onlinecite{Aronov1,Aronov2,Aronov3}.
Roughly, the time modulation of $\omega_c$ in the {\it dc} field
required in Refs.\ \onlinecite{Aronov1,Aronov2,Aronov3} for parametric
resonance emerges from the ``beatings'' of the responses to the two
{\it ac} signals. The nonparabolicity transforms these beatings into a
modulation of the cyclotron frequency. Although the increment,
$\Gamma$, for parametric resonance, induced by bichromatic microwave
irradiation, is proportional to the product ${\cal E}_1 {\cal E}_2$ of
the amplitudes of the two sources, while in Ref.\ \onlinecite{Aronov3}
it was proportional to the {\em first power} of the magnetic component
of the {\em ac} field, the ``bichromatic'' increment is much bigger.
As demonstrated above, the bichromatic increment is $\Gamma \sim
\omega_c\left(m c^2/E_0\right) \left( {\cal E}_1 {\cal
    E}_2/B^2\right)$, which should be compared to the increment
$\Gamma \sim \omega_c\left({\cal E}/B\right)$ of Ref.\ 
\onlinecite{Aronov3}.  The ratio contains a small factor $\left( {\cal
    E}/B \right)$ which is offset by the huge factor $\left( m
  c^2/E_0\right)$.

\acknowledgments

Two of the authors (MER and FvO) acknowledge the hospitality of the
Weizmann Institute of Science (supported by the Einstein Center and
LSF grant HPRI-CT-2001-00114 ) while some of this work was performed.
This work was also supported by the NSF-DAAD Collaborative Research
Grant No. 0231010 and the DFG-Schwerpunkt ``Quanten-Hall-Systeme."


\begin{references}

\bibitem{kotthaus75} J. P. Kotthaus, G. Abstreiter, J. F. Koch, 
and R. Ranvaud, Phys. Rev. Lett. {\bf 34}, 151 (1975).

\bibitem{abstreiter76} G. Abstreiter, J. P. Kotthaus, J. F. Koch, 
and G. Dorda, Phys. Rev. B {\bf 14}, 2480 (1976).

\bibitem{zudov01} M. A. Zudov, R. R. Du, J. A. Simmons, and J. L. Reno, 
Phys. Rev. B {\bf 64}, 201311 (2001).

\bibitem{mani02} R. Mani, J. H. Smet, K. von Klitzing, 
V. Narayanamurti, W. B. Johnson, and V. Umansky, Nature (London) 
{\bf 420}, 646 (2002).

\bibitem{zudov03} M.A. Zudov, R.R. Du, L.N. Pfeiffer, and K.W. West, 
Phys. Rev. Lett. {\bf 90}, 046807 (2003).

\bibitem{yang03}C. L. Yang, M. A. Zudov, T. A. Knuuttila, 
R. R. Du, L. N. Pfeiffer, and K. W. West, Phys. Rev. Lett. 
{\bf 91}, 096803 (2003).

\bibitem{mani03}R. G. Mani, J. H. Smet, K. von Klitzing, 
V. Narayanamurti, W. B. Johnson, and V. Umansky, 
preprint cond-mat/0303034.

\bibitem{dorozhkin03} S.I. Dorozhkin, Pis'ma Zh. \'{E}ksp. Teor. Fiz. 
{\bf 77}, 681 (2003). [JETP Lett. {\bf 77}, 577 (2003)].

\bibitem{willett03} R. L. Willett, L. N. Pfeiffer, K. W. West,
preprint cond-mat/0308406.

\bibitem{studenikin03} S. A. Studenikin, M. Potemski, P. T. Coleridge, 
A. Sachrajda, Z. R. Wasilewski, preprint cond-mat/0310347.

\bibitem{zudov04} M. A. Zudov, Phys. Rev. B {\bf 69}, 041304 (R) (2004).


\bibitem{mani04} R. G. Mani, V. Narayanamurti, K. von Klitzing, J. H. Smet, W. B. Johnson, and V. Umansky
Phys. Rev. B {\bf 69}, 161306 (2004).

\bibitem{mani'04}  R. G. Mani, J. H. Smet, K. von Klitzing, V. Narayanamurti, W. B. Johnson, and V. Umansky,
Phys. Rev. Lett. {\bf 92}, 146801 (2004).

\bibitem{studenikin04} S. A. Studenikin, M. Potemski, A. Sachrajda, M. Hilke, 
L. N. Pfeiffer, and K. W. West, preprint cond-mat/0404411.

\bibitem{andreev03} A. V. Andreev, I. L. Aleiner, and A. J. Millis, 
Phys. Rev. Lett. {\bf 91}, 056803 (2003).

\bibitem{bergeret03} Bergeret, B. Huckestein, and A. F. Volkov, 
Phys. Rev. B {\bf 67}, 241303 (2003).

\bibitem{shi03} J. Shi and X.C. Xie, Phys. Rev. Lett. {\bf 91}, 
086801 (2003).

\bibitem{durst03} A. C. Durst, S. Sachdev, N. Read, and 
S. M. Girvin, Phys. Rev. Lett. {\bf 91}, 086803 (2003).        

\bibitem{lei03} X. L. Lei and S. Y. Liu, 
Phys. Rev. Lett. {\bf 91}, 226805 (2003).

\bibitem{ryzhii03} V. Ryzhii, Phys. Rev. B {\bf 68}, 193402 (2003).

\bibitem{ryzhii03'} V. Ryzhii and V. Vyurkov
       Phys. Rev. B {\bf 68}, 165406 (2003).

\bibitem{dmitriev03} I. A. Dmitriev, A. D. Mirlin, and D. G. Polyakov, 
Phys. Rev. Lett. {\bf 91}, 226802 (2003).

\bibitem{vavilov04} M. G. Vavilov and I. L. Aleiner, 
Phys. Rev. B {\bf 69}, 035303 (2004).

\bibitem{dmitriev03'}I. A. Dmitriev, M. G. Vavilov, A. D. Mirlin, 
D. G. Polyakov, and I. L. Aleiner, preprint  cond-mat/0310668.

\bibitem{ryzhii70} V. I. Ryzhii, Sov. Phys. Solid State {\bf 11}, 
2078 (1970).

\bibitem{ryzhii86} V. I. Ryzhii, R. A. Suris, and B. S. Shchamkhalova, 
Sov. Phys. Semicond. {\bf 20}, 1299 (1986).

\bibitem{koulakov03} A. A. Koulakov and M. E. Raikh, 
Phys. Rev. B {\bf 68}, 115324 (2003). 

\bibitem{Kovalev04} In all experimental
  papers,\cite{mani02,zudov03,yang03,mani03,dorozhkin03,willett03,studenikin03,zudov04}
  magnetic fields were weak so that a classical description might be
  adequate. An experimental study of the effect of microwave
  irradiation on the quantum oscillations of the magnetoresistance,
  performed in stronger magnetic fields, was recently reported in
  A.~E.~Kovalev, S.~A.~Zvyagin, C.~R.~Bowers, J.~L.~Reno, and
  J.~A.~Simmons, Solid State Commun. {\bf 130}, 379 (2004).

\bibitem{Aronov1} 
I.~E.~Aronov, E.~A.~Kaner, and A.~A.~Slutskin,
  Solid. State Commun. {\bf 38}, 245 (1981).

\bibitem{Aronov2}
I.~E.~Aronov and E.~A.~Kaner, JETP Lett. {\bf 34}, 325 (1981).

\bibitem{Aronov3}
I.~E.~Aronov and O.~N.~Baranetz, Phys. Rep. {\bf 197}, 99 (1990).

\end{references}
\end{document}